\documentclass[aps,prb,twocolumn,superscriptaddress,showpacs]{revtex4-1}
\usepackage{graphicx,psfrag}
\graphicspath{{/home/user/fig/publi/muon/}}
\usepackage{amsmath}
\begin{document}


\title{Magnetic field distribution and characteristic fields of the vortex lattice for
a clean superconducting niobium sample in an external field applied along a three-fold axis
}

\author {A.~Yaouanc}
\affiliation{Institut Nanosciences et Cryog\'enie, SPSMS,
CEA and Universit\'e Joseph Fourier, F-38054 Grenoble, France}
\author{A.~Maisuradze}
\affiliation{Physik-Institut der Universit\"at Z\"urich, Winterthuerstrasse 190,
CH-8057 Z\"urich, Switzerland}
\affiliation{Laboratory for Muon-Spin Spectroscopy,
Paul Scherrer Institute, CH-5232 Villigen-PSI, Switzerland}
\author{N.~Nakai}
\affiliation{Department of Physics, Okayama University, Okayama 700-8530, Japan}
\author{K.~Machida}
\affiliation{Department of Physics, Okayama University, Okayama 700-8530, Japan}
\author{R.~Khasanov}
\affiliation{Laboratory for Muon-Spin Spectroscopy,
Paul Scherrer Institute, CH-5232 Villigen-PSI, Switzerland}
\author{A.~Amato}
\affiliation{Laboratory for Muon-Spin Spectroscopy,
Paul Scherrer Institute, CH-5232 Villigen-PSI, Switzerland}
\author{P.~K.~Biswas}
\affiliation{Laboratory for Muon-Spin Spectroscopy,
Paul Scherrer Institute, CH-5232 Villigen-PSI, Switzerland}
\author{C.~Baines}
\affiliation{Laboratory for Muon-Spin Spectroscopy,
Paul Scherrer Institute, CH-5232 Villigen-PSI, Switzerland}
\author{D.~Herlach}
\affiliation{Laboratory for Muon-Spin Spectroscopy,
Paul Scherrer Institute, CH-5232 Villigen-PSI, Switzerland}
\author{R.~Henes}
\affiliation{Max-Planck-Institut f\"ur Metallforschung,
Heisenbergstra{\ss}e 3, D-70569 Stuttgart, Germany}
\author{P.~Keppler}
\affiliation{Max-Planck-Institut f\"ur Metallforschung,
Heisenbergstra{\ss}e 3, D-70569 Stuttgart, Germany}
\author{H.~Keller}
\affiliation{Physik-Institut der Universit\"at Z\"urich, Winterthuerstrasse 190,
CH-8057 Z\"urich, Switzerland}





\date{\today}

\begin{abstract}

The field distribution in the vortex lattice of a pure niobium single crystal with an
external field applied along a three-fold axis has been investigated by the
transverse-field muon-spin-rotation (TF-$\mu$SR) technique over a wide range of temperatures
and fields. The experimental data have been analyzed with the Delrieu's
solution for the form factor supplemented by phenomenological formulas for the
parameters. This has enabled us to experimentally establish the temperatures and fields
for the Delrieu's, Ginzburg-Landau's, and Klein's regions of the vortex
lattice.  Using the numerical solution of the quasiclassical Eilenberger's equation 
the experimental results have been reasonably understood. They should apply to all clean BCS 
superconductors. The analytical Delrieu's model supplemented by phenomenological
formulas for its parameters is found to be reliable for analyzing TF-$\mu$SR experimental
data for a substantial part of the mixed phase. The Abrikosov's limit is contained in it.

\end{abstract}


\maketitle

\section{Introduction}
\label{Introduction}

The physical properties of superconductors are usually described by the
phenomenological Ginzburg-Landau (GL) theory.\cite{Tinkham96,Ketterson99}
For a type-II superconductor it predicts a mixed phase with a periodic
variation of the magnetic induction in the form of a vortex lattice (VL),
as first derived by Abrikosov.\cite{Abrikosov57}
For a simple superconductor the VL is characterized by three fields:
the minimum field, the field at the saddle point of the field map, and the
maximum field which is found in the vortex cores. The minimum field lies at
the centre of the equilateral triangle formed by three nearest neighbor vortex
cores. This simple picture is believed to be valid, although the GL's theory is
theoretically justified only in the vicinity of $T_{\rm c0}$,\cite{Gorkov59}
the critical temperature at low field. However, because of its simplicity, it
serves as a basis for data analysis of experiments performed in the whole
mixed phase.\cite{Sonier00,Sonier07} 

Using an approximate solution of the microscopic BCS-Gor'kov's equation,\cite{Gorkov59}
Delrieu discovered the minimum field in the vicinity of the upper critical
field $B_{\rm c2}$ at low temperature to be at the  midpoint between two vortex
cores.\cite{Delrieu72} Later on, solving numerically the Eilenberger's
equation \cite{Eilenberger68}  --- an analytical approximation to the
BCS-Gor'kov's equation involving an integration over the magnitude of the
electron wave vector --- Klein found two field minima
in the VL unit cell at intermediate temperature.\cite{Klein87}
Consistent with Klein's results, nuclear magnetic resonance (NMR) experiments 
by Kung \cite{Kung70} on vanadium have detected a linear temperature dependence of 
the vortex-core field in a large temperature range towards zero temperature.

Here we present an exhaustive study of the field distribution in the VL
of a pure niobium single crystal with the magnetic field ${\bf B}_{\rm ext}$
applied along a crystallographic $\langle 111 \rangle$ direction. The measurements
have been done using the transverse-field positive-muon-rotation (TF-$\mu$SR)
technique.\cite{Dalmas97,Yaouanc11} We have recently published a report which focuses on
the $B_{\rm c2}$ vicinity.\cite{Maisuradze13b} The present data have been analyzed
with an expression for the form factor derived analytically by Delrieu. Notice that the 
form factor of Abrikosov is a limiting case of the former expression. Combining experimental 
and theoretical results, we have established the VL characteristics predicted by Abrikosov, 
Delrieu and Klein in the proper parameter ranges. Our findings have been explained 
semi quantitatively using results obtained by solving numerically the Eilenberger's equation
assuming a cylindrical Fermi surface.

The organization of this paper is as follows.  Section~\ref{Theory} introduces theoretical
models for the VL field distribution. In Sec.~\ref{Experimental} the
sample is described, as well as the experimental conditions and the data analysis.
Section~\ref{Typical} displays typically measured field distributions. The following section
(Sec.~\ref{Results}) discusses the VL characteristics derived from the present experimental 
and theoretical studies. We summarize the results obtained in this work in 
Section~\ref{Summary_study}. Possible improvements of the data analysis and experimental 
conditions are mentioned. Some conclusions and perspectives are presented in Sec.~\ref{Conclusions}.

The reader only interested by the characteristics of the VL for niobium derived
from our measurements and their analysis will jump directly to Sec.~\ref{Results}.
We stress that they are expected to be found for all clean BCS superconductors.

\section{Theoretical background on the field distribution in the VL of  niobium}
\label{Theory}

We shall first shortly describe the theories used to fit the $\mu$SR asymmetry time
spectra. Then we shall present some computed field distributions.

We recall that a conventional triangular VL is observed when ${\bf B}_{\rm ext}$ is
applied along a three-fold axis as revealed by small-angle neutron scattering
(SANS).\cite{Schelten71,Kahn73,Forgan02,Muhlbauer09} In contrast to expectation,
the VL field distribution is not described by the GL theory for $B_{\rm ext}$ in the
$B_{\rm c2}$ vicinity at low temperature.\cite{Herlach90,Maisuradze13b} On the other
hand, the approximate Delrieu's solution of the BCS-Gor'kov equation explains the
measured distribution, but only above 0.6~K.\cite{Maisuradze13b} This is a temperature
much lower than the zero field critical temperature  $T_{\rm c0} = 9.25$~K.

\subsection{Description of the available theories}

\begin{figure}
\includegraphics[scale=0.70]{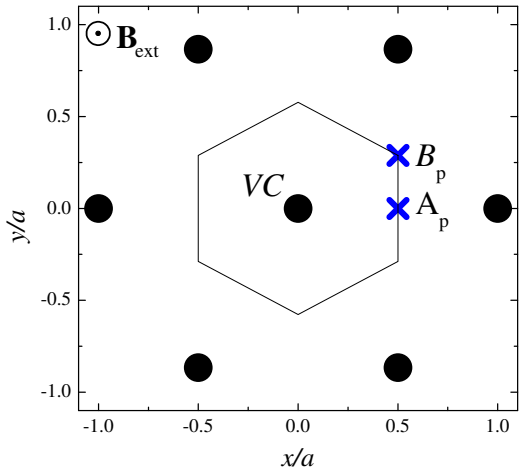}
\caption{(color online).
A triangular vortex lattice in the direct space. The external field ${\bf B}_{\rm ext}$
is applied perpendicular to the vortex plane. The bullets illustrate vortex core
coordinates and the crosses two positions of particular interest in a unit cell,
$A_{\rm p}$ and $B_{\rm p}$. In addition, we specify the position of one of the vortex core 
center with $VC$ nearby. The thin solid line represents the Wigner-Seitz cell.
The position coordinates are scaled by the VL parameter $a$.
}
\label{geometry}
\end{figure}

We illustrate in Fig.~\ref{geometry} an equilateral triangular VL.
Three positions of special interest have been defined: the midpoint between two consecutive
vortex cores, i.e.\ $A_{\rm p}$, and the centre of an equilateral triangle formed by
three nearest neighbor vortex cores, i.e.\ $B_{\rm p}$. In addition, a vortex-core position
is labeled as $VC$.

The TF-$\mu$SR technique gives access to the component field distribution along the
${\bf B}_{\rm ext}$ direction if $B_{\rm ext}$ is sufficiently
large.\cite{Yaouanc11} This is the case here. Labeling this direction as $Z$,
${\tilde D}_{\rm c}(B^Z)$ is measured. It can be computed from the real space field map
$B^Z({\bf r})$ of the two-dimensional VL:
\begin{eqnarray}
{\tilde D}_{\rm c}(B^Z) = \int_{\rm u.c.} \delta (B^Z({\bf r}) -B^Z) {\rm d}^2 {\bf r},
\label{Theory_2}
\end{eqnarray}
where the integral extends over the VL unit cell. In terms of its Fourier
components $B^Z_{{\bf K}_{m,n}}$, we have for the field map
\begin{eqnarray}
B^Z({\bf r}) = \sum_{{\bf K}_{m,n}} B^Z_{{\bf K}_{m,n}}
\exp(i{{\bf K}_{m,n}} \cdot {\bf r} ),
\label{Theory_21}
\end{eqnarray}
where the sum is over the reciprocal space of the VL.

We need an expression for $B^Z_{{\bf K}_{m,n}}$ which is usually called the form
factor in the SANS literature.

First of all we note that our sample is in the clean limit.\cite{Maisuradze13b} So we
do not need to consider the effect of impurities. We shall use two theories for
the computation of $B^Z_{{\bf K}_{m,n}}$ that are approximations to the BCS-Gor'kov theory.
Two types of approximations are to be considered: the ones common and the ones particular
to a theory. Common is the semiclassical approximation.
Here the hypothesis is made that the spacing between the Landau levels is small in
comparison to the sum of their thermal and collision broadenings.\cite{Werthamer69} This
is expected to be valid for niobium down to $\approx 0.04$~K.\cite{Maisuradze13a} To get
quantitative predictions to compare with experimental data, we need a simple, but still
realistic, Fermi surface. The Delrieu's solution of the BCS-Gor'kov equation assumes
a spherical Fermi surface. For an extensive numerical study of the  Eilenberger's equation
such as presented here, a cylindrical Fermi surface is a natural choice. This is partly 
because of computational convenience and is believed to be enough for the present purpose 
to constructing the phase diagram. Its essential features and other 
properties in this paper do not change when a 3-dimensional Fermi surface model is used,
merely changing the $\kappa_{\rm GL}$ value. We can translate and interpret the $\kappa_{\rm GL}$ 
values between the 2-dimensional and 3-dimensional cases. As for more realistic Fermi surface models, 
some of us have an experience to use a realistic 3-dimensional Fermi surface model calculated 
by band theory for niobium.\cite{Adachi11} In order to evaluate the subtle vortex lattice 
orientational changes for ${\bf B}_{\rm ext} \parallel \langle 001 \rangle $ it is definitely 
needed to have a realistic Fermi surface model. For the present purposes the cylindrical Fermi 
surface model is believed to be enough.

Next, for completeness we summarize the work of Delrieu. He neglected the spatial dependence of the order 
parameter $\Delta({\bf r})$. While this approximation is reasonable in the vicinity of $B_{\rm c2}$, 
it should break down when approaching the lower critical field. He
derived $B^Z_{{\bf K}_{m,n}} = f_{m,h}({\tilde a}, {\tilde b}, {\tilde c})$.
The function $f_{m,h}$  can be found
elsewhere.\cite{Delrieu72,Maisuradze13a} Here ${\tilde a}$ has the dimension
of a field:
\begin{eqnarray}
{\tilde a} = -{\mu_0 N_0 \Delta^2_0 \over 2 {\overline {B^Z}}} {\tilde c}.
\label{Theory_51}
\end{eqnarray}
In the region of validity of the Delrieu's approximation,
$\overline{B^Z ({\bf r})} \equiv \overline{B^Z } \simeq B_{\rm ext}$.
The parameter ${\tilde a}$ does not influence the
shape of $B^Z_{{\bf K}_{m,n}}$ and therefore ${\tilde D}_{\rm c}(B^Z)$. It only gives its scale.
It is proportional to the density of state at the Fermi level in the normal metal $N_0$
(per spin, volume, and energy), the quantity $\Delta^2_0 = \overline{|\Delta({\bf r})|^2}$
($\overline{|\Delta({\bf r})|^2}$ is the spacial average of $|\Delta({\bf r})|^2$), and is
inversely proportional to the average field $\overline{B^Z}$.
The dimensionless parameters ${\tilde b}$ and ${\tilde c}$ determine the ${\tilde D}_{\rm c}(B^Z)$
shape and are expressed in terms of the ratios of three length scales:
${\tilde b} = [\Lambda/(\pi \xi^B)]^2$ and ${\tilde c} = \Lambda/\xi^T$. Here,
$\Lambda = [\Phi_0/(2 \pi \overline{B^z}]^{1/2}$ is a length parameter proportional to the
intervortex distance. The field and temperature dependent length scale
$\xi^B = \hbar v_{\rm F}/(\pi \Delta_0)$ diverges near $B_{\rm c2}$, while
$\xi^T = \hbar v_{\rm F}/(2 \pi k_{\rm B} T)$. We have introduced the Fermi velocity
$v_{\rm F}$. It is easily found that
\begin{eqnarray}
{\tilde b} = {1 \over \pi^2} {\xi^2_{\rm GL} (T) \over \xi^2_0(T)} {1 -b \over b},
\label{Theory_3}
\end{eqnarray}
where $b = B_{\rm ext}/B_{c2}(T)$ is the reduced field, $\xi_{\rm GL}$ the GL coherence length,
and $\xi_0$ Pippard-BCS coherence
length.\cite{Maisuradze13a} To derive Eq.~\ref{Theory_3} two phenomelogical formulas
expected to be valid for conventional superconductors have been used:
\begin{eqnarray}
\Delta_0 = \Delta_0(0) \sqrt{ 1-b} \sqrt{1 - \tau^2},
\label{Theory_4}
\end{eqnarray}
and
\begin{eqnarray}
B_{\rm c2}(T) = B_{\rm c2}(0) (1 - \tau^2),
\label{Theory_5}
\end{eqnarray}
where $\tau = T/T_{\rm c0}$.
Since ${\xi^2_{\rm GL} (T) \over \xi^2_0(T)} = {\xi^2_{\rm GL} (0) \over \xi^2_0(0)}$
and ${\xi_{\rm GL} (0) \simeq \xi_0(0)/0.96}$ in the clean limit,\cite{Tinkham96}
\begin{eqnarray}
{\tilde b} = 0.110 {1 -b \over b}.
\label{Theory_6}
\end{eqnarray}
Hence, ${\tilde b}$ only depends on $b$.
The parameter ${\tilde c}$ can be expressed in terms of $v_{\rm F}$:\cite{Maisuradze13a}
\begin{eqnarray}
{\tilde c} =
{\sqrt {\Phi_0  2 \pi} k_{\rm B} T \over \sqrt{\overline {B^Z}} \hbar v_{\rm F} },
\label{Theory_7}
\end{eqnarray}
where $\Phi_0$ is the magnetic flux quantum.
Interestingly, $N_0 \Delta_0^2$ in Eq.~\ref{Theory_51} is the condensation
energy $E_{\rm c}$.\cite{Maisuradze13b} Hence, according to Eq.~\ref{Theory_4},
\begin{eqnarray}
E_{\rm c} = E_{\rm c}(0) ( 1-b) (1 - \tau^2).
\label{Theory_8}
\end{eqnarray}
This means that $B^Z_{{\bf K}_{m,n}}$ from the Delrieu's solution supplemented
by phenomelogical formulas for the parameters depends on two material parameters:
$v_{\rm F}$ and $E_{\rm c}(0)$, the second parameter being only involved in the
scaling of the field. Introducing the unitless field
\begin{eqnarray}
b_{\rm N}^Z = {B^Z - B_{\rm sad} \over B_{\rm vc} - B_{\rm sad}},
\label{Theory_9}
\end{eqnarray}
where $B_{\rm sad}$ and $B_{\rm vc}$ are the saddle point and vortex core fields discussed
at length in Sec~\ref{Comparison}. While $B_{\rm vc}$ is always the field at a vortex-core center, the 
position of $B_{\rm sad}$ may change as described in Sec. \ref{Results_fields}. The unitless component 
field distribution $D_{\rm c}(b^Z_{\rm N})$ only depends
on $v_{\rm F}$. However, for the computation of a measured TF-$\mu$SR asymmetry time spectrum
we need  ${\tilde D}_{\rm c}(B^Z)$ rather than $D_{\rm c}(b^Z_{\rm N})$. Since
$D_{\rm c}(b^Z_{\rm N}) {\rm d}b^Z_{\rm N}   = {\tilde D}_{\rm c}(B^Z) {\rm d} B^Z$,
\begin{eqnarray}
{\tilde D}_{\rm c}(B^Z)   = {1 \over B_{\rm vc} - B_{\rm sad}}D_{\rm c}(b^Z_{\rm N}).
\label{Theory_10}
\end{eqnarray}
Hence, as expected, ${\tilde D}_{\rm c}(B^Z)$ depends on two materials parameters, namely $v_{\rm F}$
and $E_{\rm c}(0)$, and $D_{\rm c}(b^Z_{\rm N})$ only on $v_{\rm F}$.

Since the  $B^Z_{{\bf K}_{m,n}}$ analytical Delrieu's solution derives from an
approximation to the BCS-Gor'kov theory supposed to be valid only in the $B_{\rm c2}$
vicinity, we need a method to compute  $B^Z_{{\bf K}_{m,n}}$ for the whole VL.
In addition, if possible, it would be nice not to rely on phenomenological
formulas for the physical parameters. The Eilenberger's equation for the thermal
Green's functions fit our purpose. Eilenberger introduced Green's functions that
result from the Gor'kov's functions integrated over the magnitude of the electron
wave vector.\cite{Eilenberger68} These former functions follow transport-like equations suitable 
for numerical calculations as first shown by Klein.\cite{Klein87} Supplemented by the self-consistent 
equations for the gap function and vector potential,  here we have directly computed $B^Z({\bf r})$ 
normalized by $B_{\rm c2}(0)$, a quantity directly observable. 

The integration over the magnitude of the wave vector introduces an approximation
which is valid when $k_{\rm F} \xi_0 \gg 1$, where $k_{\rm F}$ is the Fermi wave vector.
Since $1/k_{\rm F}$ is of the order of the niobium lattice parameter and
$\xi_0 \simeq 27$nm,\cite{Maisuradze13a} the condition $k_{\rm F} \xi_0 \gg 1$
is clearly fulfilled. Following Brandt's method for solving the GL's
equations,\cite{Brandt97} the Eilenberger's equation are nowadays solved taking
advantage of the periodicity of the VL.\cite{Miranovic03} Nicely enough,
$B^Z({\bf r})/B_{\rm c2}(0)$ depends only on one single material parameter: the GL parameter
$\kappa_{\rm GL} = \lambda/\xi_{\rm GL}$, where $\lambda$ is the london penetration
depth. This is to be compared to $B^Z_{{\bf K}_{m,n}}/{\tilde a}$ from Delrieu
which also depends on one single parameter, but $v_{\rm F}$ rather than $\kappa_{\rm GL}$.
These two parameters are related.\cite{Klein87}

\subsection{Characteristics of field distributions}
\label{Comparison}

\begin{figure}
\includegraphics[scale=0.50]{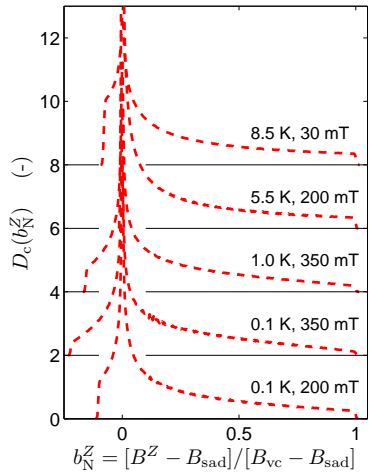}
\caption{(color online).
Component field distributions $D_{\rm c}(b^Z_{\rm N})$ computed with the Delrieu's model
assuming $v_{\rm F} = 2.0 \times 10^5$~m/s. Five cases are considered: from top to bottom, in 
the $B_{\rm c2}(T)$ vicinity at four temperatures, i.e.\ $T= 8.5$, $5.5$, $1.0$ and $0.1$~K, 
and at low temperature and field but still in the mixed phase, i.e.\ $T = 0.1$~K with 
$B_{\rm ext} = 200$~mT.  
}
\label{Theory_fig}
\end{figure}
We compare $D_{\rm c}(b^Z_{\rm N})$ for some selected $(T,B_{\rm ext})$ values computed from 
the analytical Delrieu's theory in Fig.~\ref{Theory_fig}. We take the material parameter 
valid for our niobium sample:\cite{Maisuradze13b} $v_{\rm F} = 2.0 \times 10^5$~m/s.  We recall that 
unitless $D_{\rm c}(b^Z_{\rm N})$ only depends on $v_{\rm F}$. We recall that $T_{\rm c0} = 9.25$~K and
$B_{\rm c2} (0) = 0.43$~T. As expected and clearly seen in Fig.~\ref{Theory_fig}, a
distribution is characterized by three fields: its minimum field  $B_{\rm min}$, a saddle point
field $B_{\rm sad}$ in the field map for which $D_{\rm c}(b^Z_{\rm N})$ displays a maximum,
and $B_{\rm vc}$ which is the field in the center of a vortex core, i.e.\ the maximum field in
$D_{\rm c}(b^Z_{\rm N} )$. Note that two features of a
distribution are strongly dependent on the $(T,B_{\rm ext})$ values: the shape of the
high-field tail and the distance between $B_{\rm min}$ and $B_{\rm sad}$.
As shown previously,\cite{Maisuradze13b} the observation of a linear high-field
tail for large $B_{\rm ext}$ and low $T$ --- clearly seen for the fourth distribution from
the top --- is a signature in those experimental conditions of the pronounced conical shape
of the field variation around the vortex cores. This results from the partial Cooper pair
diffraction on the vortex cores. We find it convenient to measure the distance between
$B_{\rm min}$ and $B_{\rm sad}$ with the following unitless normalized ratio:
\begin{eqnarray}
\delta B^{\rm n}_{\rm sad,min} = -{B_{\rm min} - B_{\rm sad} \over B_{\rm vc} - B_{\rm sad}}.
\label{Theory_12}
\end{eqnarray}
The first three distributions from the top of Fig.~\ref{Theory_fig} concern $D_{\rm c}(b^Z_{\rm N})$
with $B_{\rm ext}$ in the $B_{\rm c2}(T)$ vicinity. A  $\delta B^{\rm n}_{\rm sad,min}$ minimum
is predicted  around $T=5.5$~K. This should easily be observed experimentally. Although,
because of the Gaussian smearing discussed in the next section (Sec.~\ref{Experimental}),
$\delta B^{\rm n}_{\rm sad,min}$ is not expected to be as small as predicted. We postpone the
discussion of its physical meaning
to Sect.~\ref{Results}. A close look at the last two $D_{\rm c}(b^Z_{\rm N})$ from the top of
Fig.~\ref{Theory_fig} illustrates the effect of the field intensity at low temperature.
An exotic  $D_{\rm c}(b^Z_{\rm N})$ is only predicted for a sufficiently large $B_{\rm ext}$.

\section{Experimental}
\label{Experimental}

Here the sample is described, as well as the experimental conditions and the
data analysis.

The TF-$\mu$SR measurements reported here have been performed on the single crystal
described in Ref.~\onlinecite{Maisuradze13b}. The small $B_{\rm c2} = 430 \,(2)$~mT
testifies of its high quality and purity, as well as the lack of difference
between the distributions measured with the zero-field-cooled or
field-cooled procedures at 1.5~K under $B_{\rm ext} = 360$~mT.

The new TF-$\mu$SR measurements described here have again been performed at the
Swiss Muon Source (S$\mu$S), Paul Scherrer Institute (PSI), Switzerland, using
the general purpose spectrometer (GPS) and Dolly spectrometers for $T \geq 1.6$~K.
Measurements for $T<1.6$~K have been conducted on the low temperature facility (LTF) 
spectrometer. 

\begin{figure}
\includegraphics[scale=0.70]{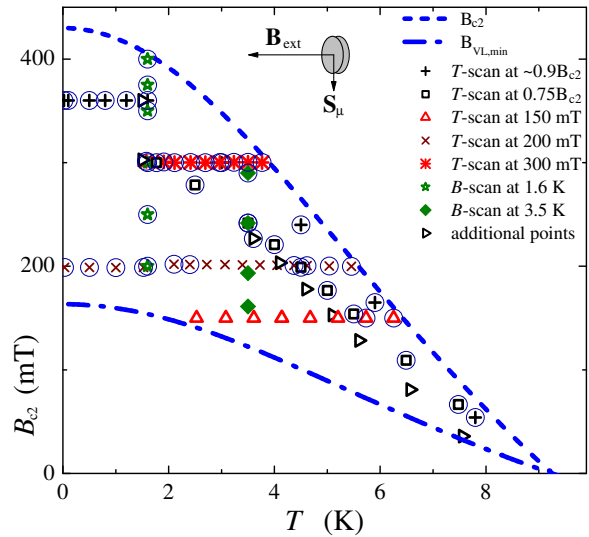}
\caption{(color online).
Temperatures and fields at which measurements have
been performed on the niobium single crystal.
The geometry for a measurement is depicted by the
pictogram: the directions of ${\bf B}_{\rm ext} \parallel \langle 111 \rangle$
and of the initial muon beam polarization ${\bf S}_\mu$ are
given. The $B_{\rm VL,min}(T)$ and $B_{\rm c2}(T)$ lines have been determined as
explained in the main text. The points at which the form factor from Delrieu
provides a proper description of the experimental $\mu$SR data are encircled.
}
\label{Measurement}
\end{figure}

\begin{figure}
\includegraphics[scale=0.50]{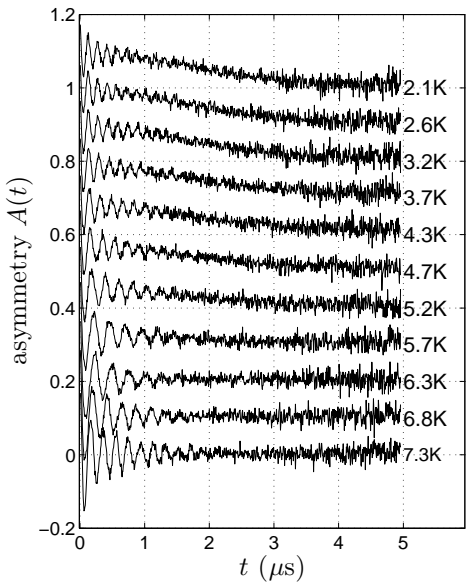}
\caption{
Some asymmetry time spectra for our niobium single crystal with
${\bf B}_{\rm ext} \parallel \langle 111 \rangle$ as a function of temperature
at $B_{\rm ext}=50$~mT.
}
\label{Measurement:500G}
\end{figure}

Our niobium sample is a single crystal disk of 13~mm diameter and 2~mm thickness
with a three-fold axis oriented normal to the disk. In Fig.~\ref{Measurement}
we specify the values of the temperatures and fields for which measurements have been
done. Five temperature scans have been performed at $B_{\rm ext}=$~300, 200 and 150~mT,
and at $0.75 B_{\rm c2}$ and $\sim 0.9 B_{\rm c2}$. In addition, we report two field
scans at 1.6 and 3.5~K. The $B_{\rm c2}(T)$ line has been determined
previously.\cite{Maisuradze13b}
In addition to the traditional VL and Meissner phases, one needs to consider
the intermediate state.\cite{Laver06} The VL phase is characterized by a single
damped oscillation centered around zero as seen for $T= 7.3$ and $6.8$~K in
Fig.~\ref{Measurement:500G}.
The amplitude of the oscillating signal at time $t=0$ is proportional to the fraction of the 
niobium sample in the VL state. When cooling down to 5.7~K the oscillation is no longer centered 
around zero and its amplitude is reduced. The signals below 5.7~K represent the sum of an oscillating 
and Kubo-Toyabe type components. This indicates that the sample has left the mixed phase and part 
of it is in a zero field condition due to Meissner screening. Performing a few series 
of measurements as reported in Fig. \ref{Measurement:500G} we have determined the $B_{\rm VL,min}(T)$ 
line shown in Fig.~\ref{Measurement}. For $B_{\rm ext} > B_{\rm VL,min}(T)$ the sample is certainly in 
the mixed phase. We shall use this conservative estimate.

The analysis of the asymmetry time spectra has been done following the method explained in
Ref.~\onlinecite{Maisuradze13b}. Here we would like to stress three points. We do fit the time
spectra and not the component field distributions which are only computed for display purpose.
Such a distribution is denoted as  ${\tilde D}^{\rm exp}_{\rm c}(B^Z)$. The difference between
${\tilde D}^{\rm exp}_{\rm c}(B^Z)$ and ${\tilde D}_{\rm c}(B^Z)$ arises from the contributions of the nuclear
$^{93}$Nb magnetic moments and the VL disorder to the field distribution at the muon site.
These contributions are taken into account in the fits by a single Gaussian
function.\cite{Brandt74} This leads to a Gaussian smearing of ${\tilde D}_{\rm c}(B^Z)$. The influence
of disorder is relatively modest in the high-field part of a distribution.\cite{Maisuradze09}
Finally, it has been shown previously that no effect of the muon diffusion
on the measured ${\tilde D}^{\rm exp}_{\rm c}(B^Z)$ is expected.\cite{Maisuradze13b}

Data analysis with the Delrieu's model is exceedingly time consuming. The
computation of ${\tilde D}_{\rm c}(B^Z)$ may take few minutes. Since a large number
of iterations are needed for fitting a single asymmetry spectrum, an
analysis would take many hours. In order to
accelerate the analysis we have first computed
${B}^Z_{{\bf K}_{m,h}}(\tilde{b}_i,\tilde{c}_j)$ setting $\tilde{a}=1$
since it is only a multiplicative factor. We have taken
$-30\leq m\leq30$ and $-30\leq h\leq30$ and a discrete set of $\tilde{b}_i$
and $\tilde{c}_j$ parameters, i.e.\
$\tilde{b}_i = 10^{n_i}$ and $\tilde{c}_j = 10^{n_j}$, with
$n_{\alpha} = -3 + 0.05\times (\alpha-1)$ ($\alpha$ stands for $i$ or $j$).
The indicies  $i = 1,\, 2,\, ...,\, 81$ and $j=1,\, 2,\, ...,\, 101 $
correspond to $0.001\leq\tilde{b}\leq 10$ and $0.001\leq\tilde{c}\leq 100$.
Because ${B}^Z_{{\bf K}_{m,h}}$ is a continuous function of its variables,
in the fitting procedure the actual values of
${B}^Z_{{\bf K}_{m,h}}(\tilde{a}, \tilde{b}, \tilde{c})$ were evaluated by
interpolation from the precalculated values of
$\tilde{a}\times{B}^Z_{{\bf K}_{m,h}}(\tilde{b}_i,\tilde{c}_j)$
(four-dimensional matrix). A quadratic interpolation has been used to avoid
zero second order derivatives during the $\chi^2$ minimization. With this
method an evaluation of asymmetry time spectrum or ${\tilde D}_{\rm c}(B^Z)$
can be performed within a fraction of second.

\section{Typical measured field distributions}
\label{Typical}

In this section typical measured field distributions $D^{\rm exp}_{\rm c}(b^Z_{\rm N})$ are
displayed. The curves result from a combined fit of the measured asymmetry time spectra to the
Delrieu's theory with $v_{\rm F}$ as a global fitting parameter. 
The parameters extracted from it are discussed in the next
section (Sec.~\ref{Results}).

\begin{figure}
\includegraphics[scale=0.31]{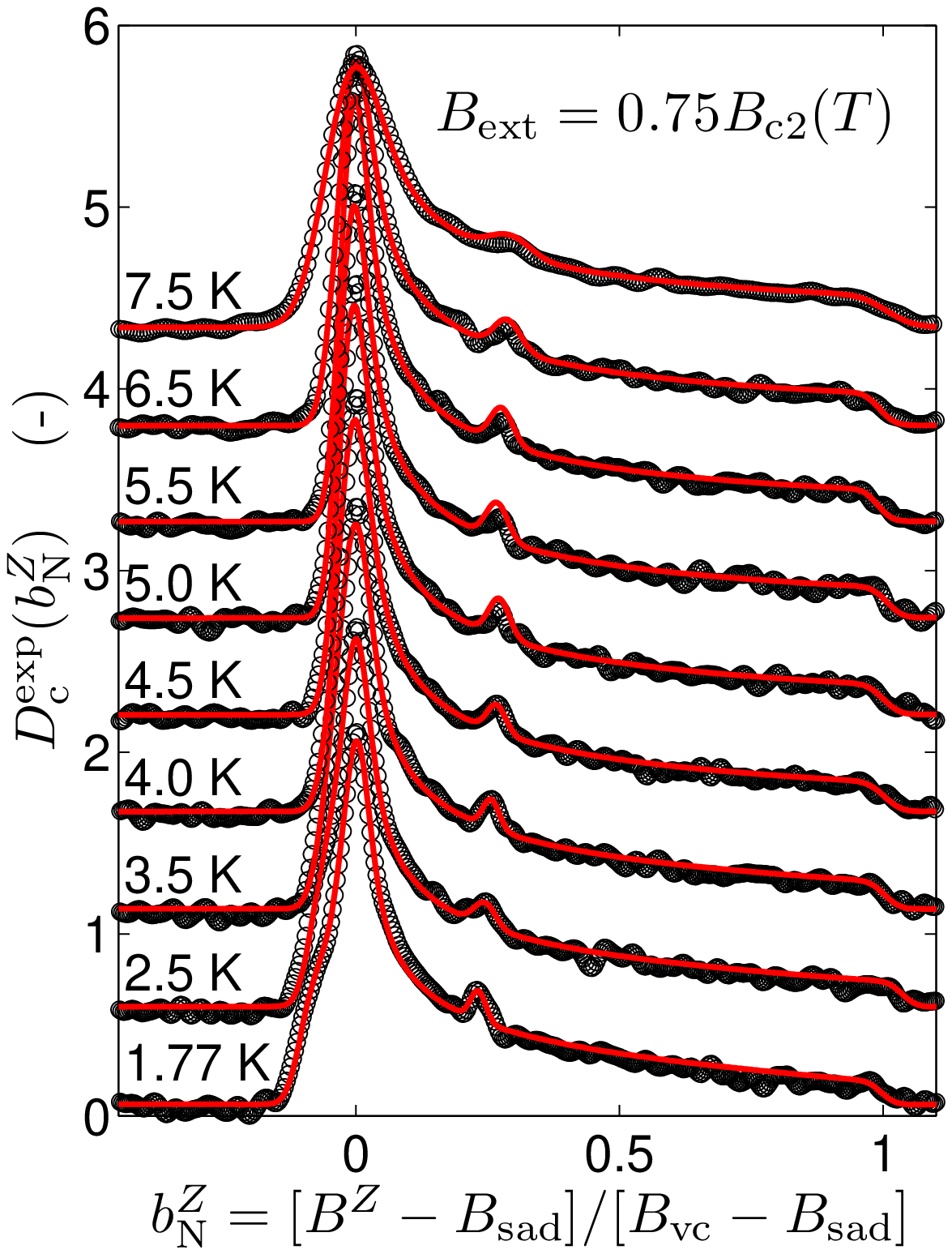}
\includegraphics[scale=0.31]{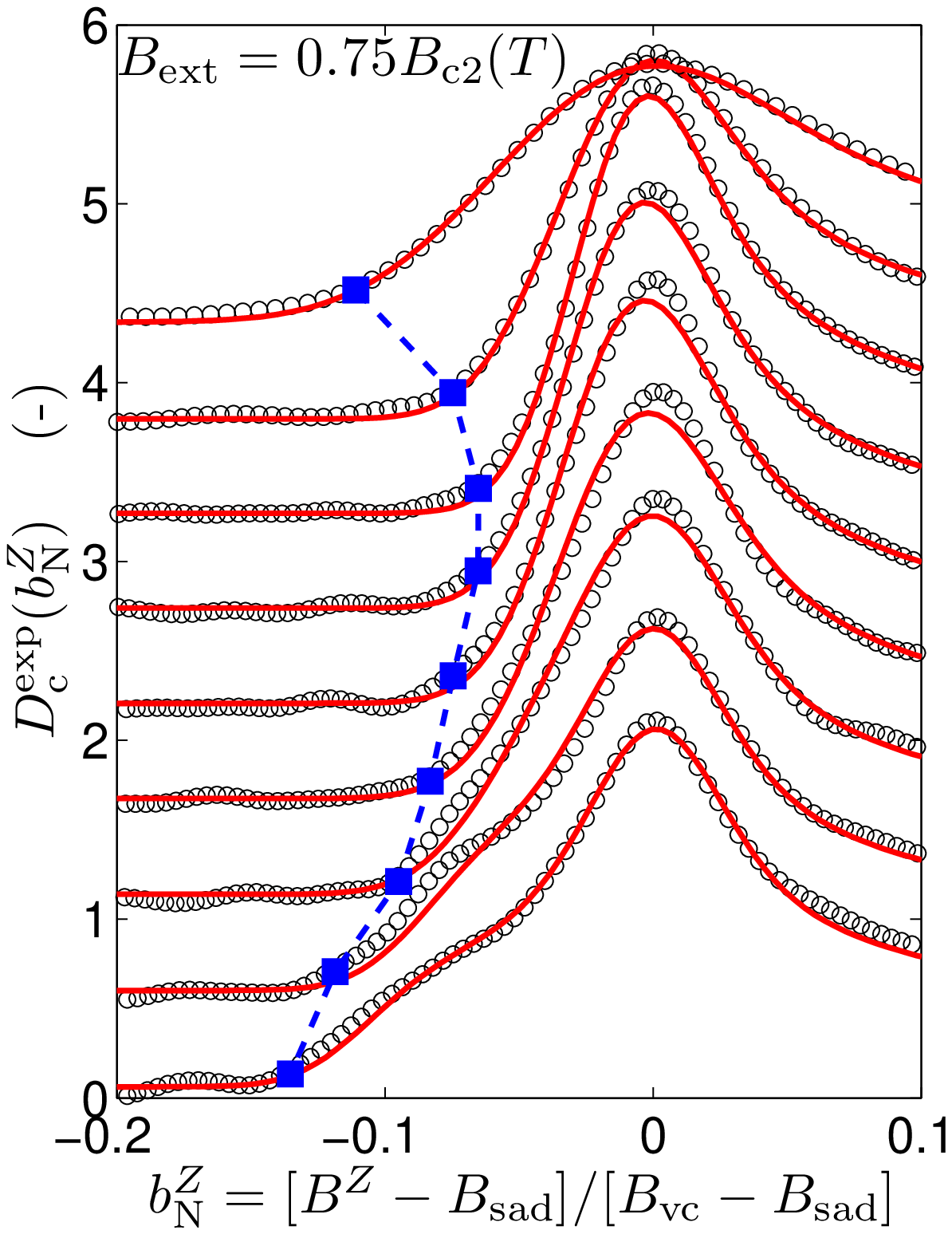}
\caption{(color online).
Some  $D_{\rm c}^{\rm exp}(b^Z_{\rm N})$ distributions as a function of temperature for
$B_{\rm ext} = 0.75 B_{\rm c2}(T)$. The data and curves have been shifted vertically for
better visualization.  For each $D_{\rm c}^{\rm exp}(b^Z_{\rm N})$ the weak intensity peak
for $b^Z_{\rm N}$ slightly larger than $0.2$ arises from the contribution of the
background --- sample holder and cryostat  walls --- to the distribution. The right panel focuses
on the low field regime. The solid lines result from fits to the Delrieu's theory as
described in Sect.~\ref{Theory}. The dashed line links the $B_{\rm min}$ values.
}
\label{fig_scan1}
\end{figure}

We start by considering the $0.75 B_{\rm c2}(T)$ temperature scan. As seen from
Fig.~\ref{Measurement}, it probes the VL from near $T_{\rm c0}$
down to low temperature, i.e.\ from 7.5 to 1.77~K. Figure~\ref{fig_scan1}
illustrates some $D^{\rm exp}_{\rm c}(b^Z_{\rm N})$. The Delrieu's theory provides a good description.
This is notified in Fig.~\ref{Measurement} by encircling the symbols which specify the
temperatures and fields of the scan. The determination of $B_{\rm min}$ at high temperature
is not very precise due to Gaussian smearing. However,
from Fig.~\ref{fig_scan1} it is quite clear that $\delta B^{\rm n}_{\rm sad,min}$ is minimum around
$5.0-5.5$~K.

Some $D^{\rm exp}_{\rm c}(b^Z_{\rm N})$ from the field
\begin{figure}
\includegraphics[scale=0.31]{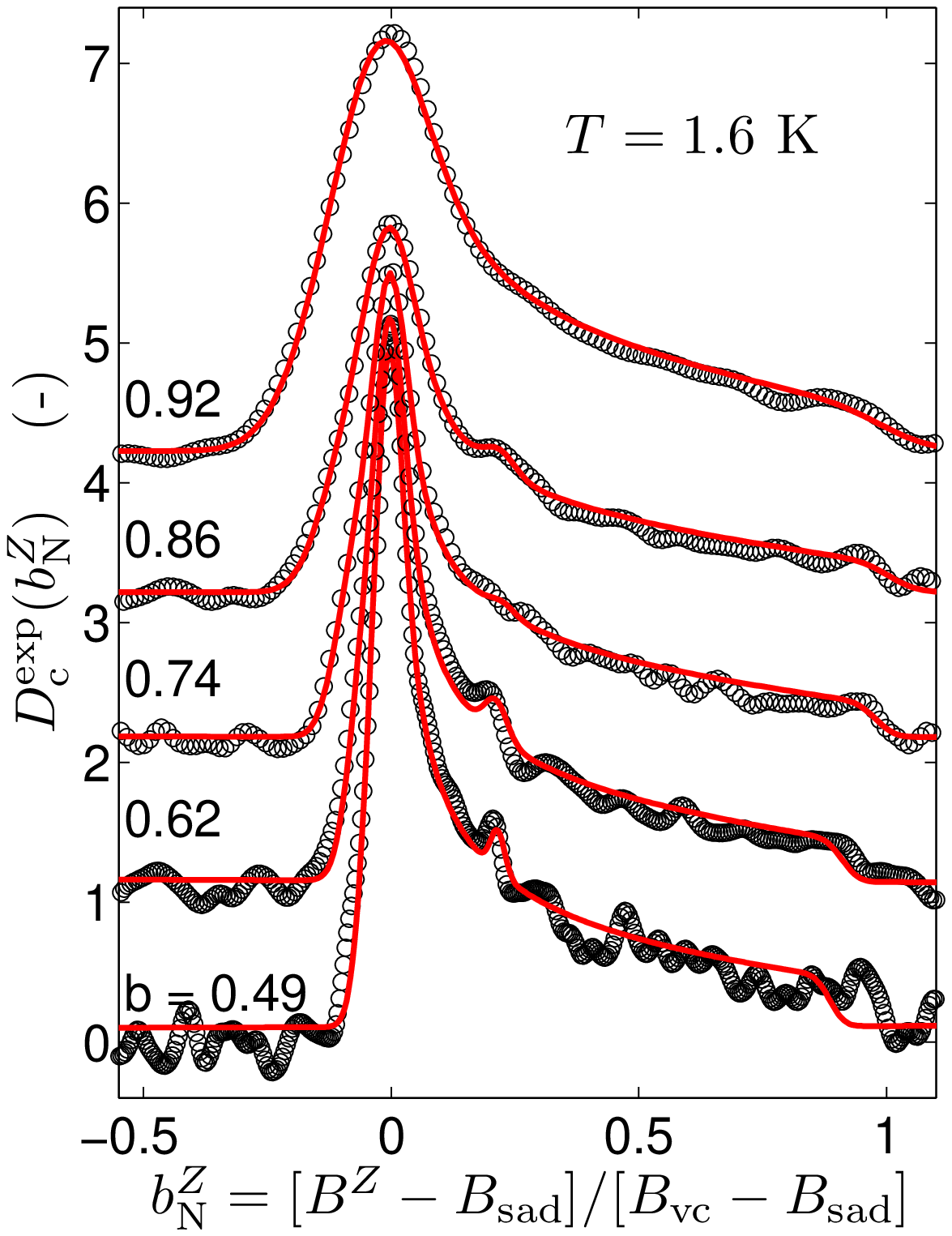}
\includegraphics[scale=0.31]{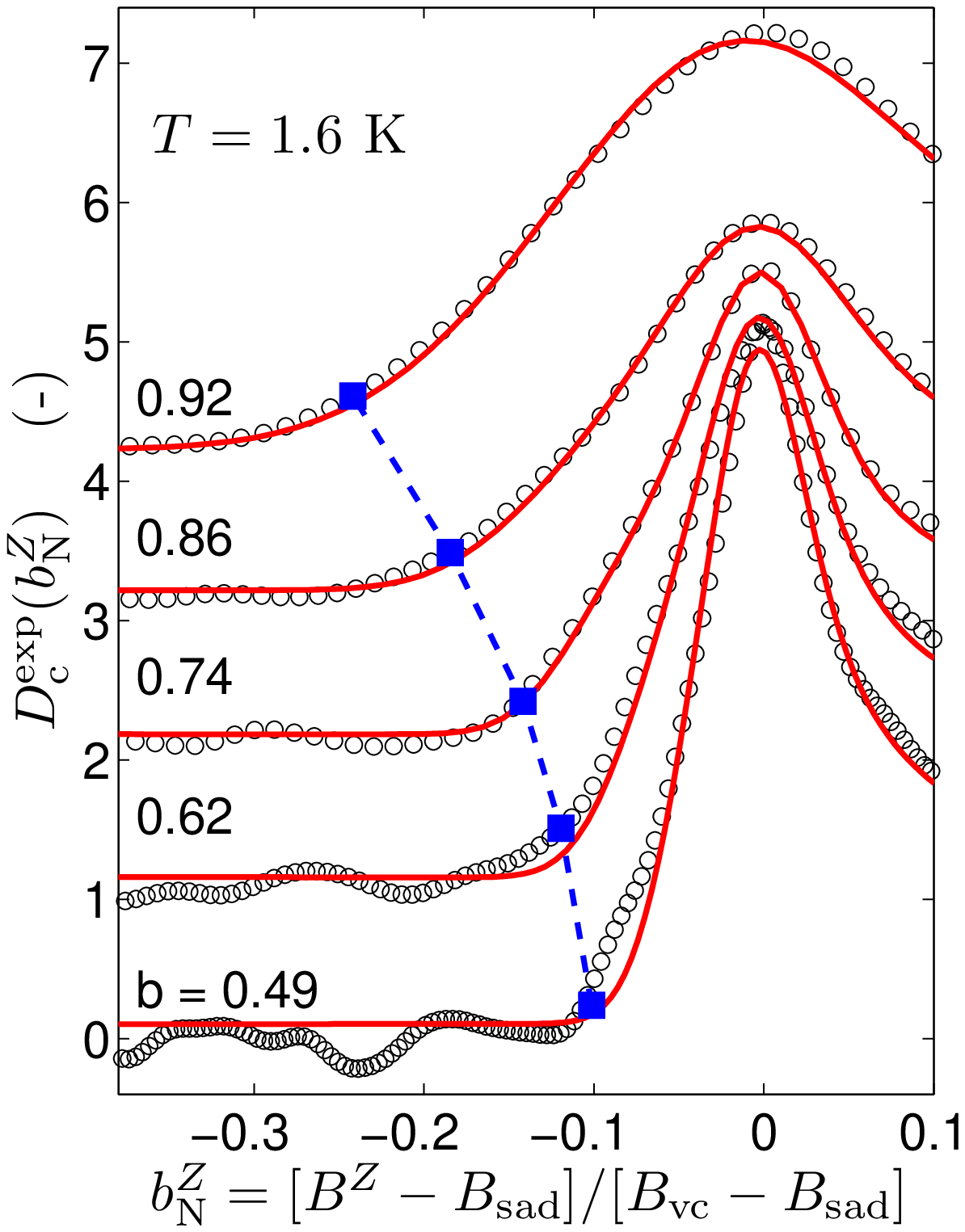}
\caption{(color online).
A selection of $D_{\rm c}^{\rm exp}(b^Z_{\rm N})$ distributions as a function of
the reduced field $b = B_{\rm ext}/B_{\rm c2}(1.6~{\rm K})$.
The data and curves have been shifted vertically for
better visualization.  For each $D_{\rm c}^{\rm exp}(b^Z_{\rm N})$
the weak intensity peak for $b^Z_{\rm N} \simeq 0.2$
arises from the contribution of the background --- sample holder and cryostat  walls ---
to the distribution.
The limited statistic for the $b = 0.49$ distribution explains its rather noisy nature.
The right panel focuses on the low field regime. The solid lines
result from fits to the Delrieu's theory as described in Sect.~\ref{Theory}. The dashed line
links the $B_{\rm min}$ values.
}
\label{fig_scan2}
\end{figure}
scan at 1.6~K are shown in Fig.~\ref{fig_scan2}. Here a smooth $\delta B^{\rm n}_{\rm sad,min}$
increase is observed as $B_{\rm c2}$ is approached. Again, Delrieu provides a good description.

\begin{figure}
\includegraphics[scale=0.31]{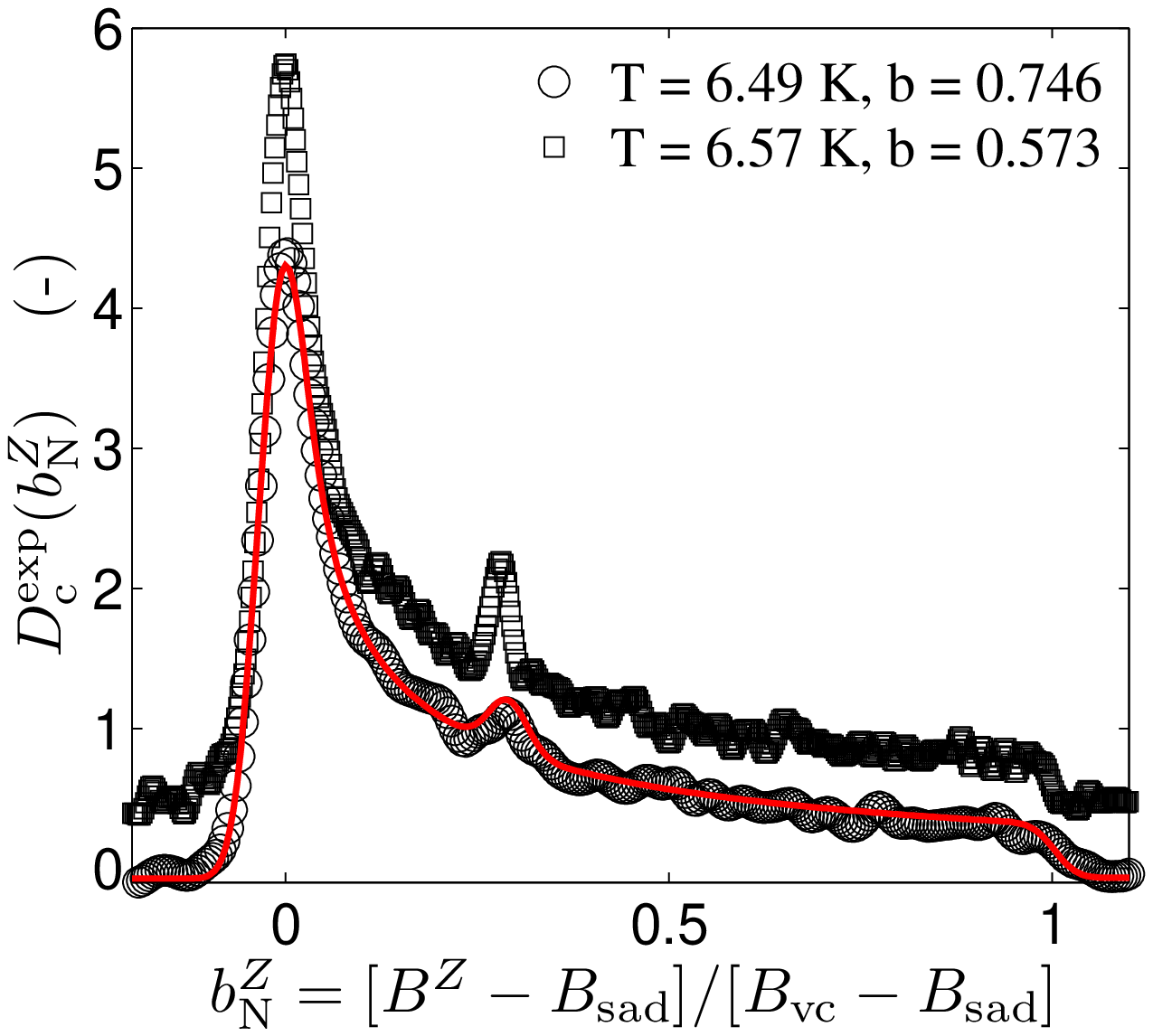}
\includegraphics[scale=0.31]{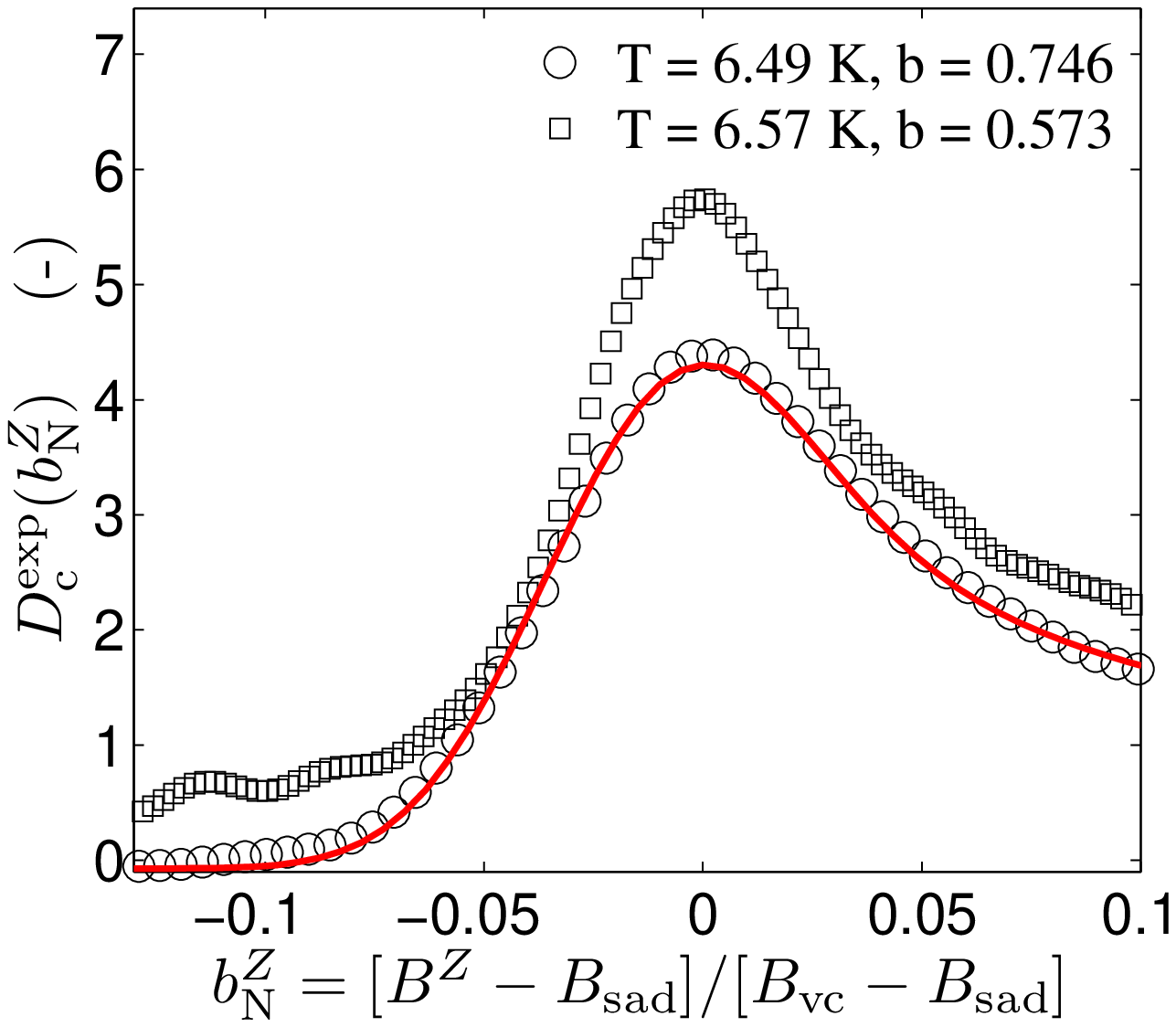}
\caption{(color online).
Two $D_{\rm c}^{\rm exp}(b^Z_{\rm N})$ distributions for $T \simeq 6.5$~K at two different fields.
The data and curves have been shifted vertically for
better visualization. The peaks around $b^Z_{\rm N} \simeq 0.3$
arise from the contributions of the background --- sample holder and cryostat  walls ---
to the distributions.
The right panel focuses on the low field regime.
The value of the reduced field $b = B_{\rm ext}/B_{\rm c2}(1.6~{\rm K})$ is indicated for
each $D_{\rm c}^{\rm exp}(b^Z_{\rm N})$. A larger contribution of the VL disorder to the
higher field distribution is seen by the
reduction of the probability at $B_{\rm sad}$.
The solid lines result from fits to the Delrieu's theory as described in Sect.~\ref{Theory}
for $T= 6.49$~K and $b=0.746$. That theory does not fit the data for $T= 6.57$~K and $b=0.573$.
}
\label{fig_scan3}
\end{figure}

Two $D^{\rm exp}_{\rm c}(b^Z_{\rm N})$ are displayed in Fig.~\ref{fig_scan3} for $T \simeq 6.5$~K.
The remarkable feature here is that $\delta B^{\rm n}_{\rm sad,min}$ is about the same
for the two distributions although the reduced field $b$ is clearly different. This is
in contrast to the two previous sets of distributions for which
$\delta B^{\rm n}_{\rm sad,min}$ is changing as a function of field or temperature.
The Delrieu's model is unable to account for the data at the lowest $B_{\rm ext}$ value.

\section{Characteristics of the mixed phase of niobium}
\label{Results}

Here the physical properties of niobium and its VL deduced from our measured $\mu$SR data are
discussed. We shall first consider the parameters extracted from the global fit
of the asymmetry time spectra with the Delrieu's approximation for the form factor.
The validity regime of the approximation will be determined. Then the properties of the
three characteristic fields of the VL will be analyzed with the numerical solution
of the Eilenberger's theory. Finally, combining our experimental results and the
Eilenberger's theory, the field map of the VL will be established.

\subsection{VL parameters  and region of validity of the Delrieu's approximation}
\label{Results_parameters}

\begin{figure}
\includegraphics[scale=0.70]{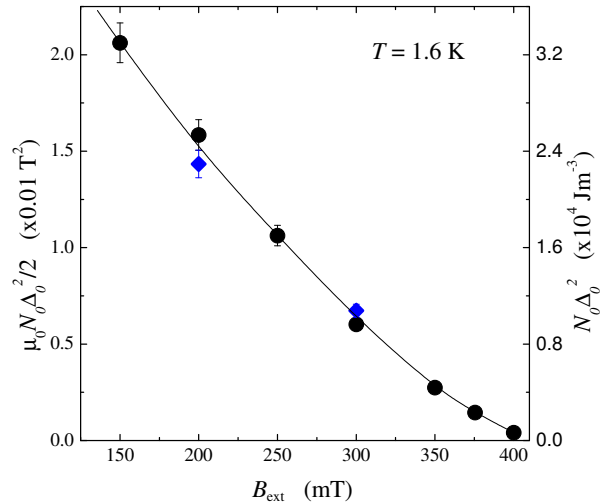}
\caption{(Color online) Condensation energy $E_{\rm c}=N_0\Delta_0^2$ and
$\mu_0 N_0\Delta_0^2/2$ as a function of $B_{\rm ext}$. The values
extracted from the field scan at $1.6$~K are represented as bullets, while
diamonds correspond to the two temperature scans at $200$ and $300$~mT.
The solid line is a guide to the eyes.
}
\label{condensation_energy}
\end{figure}

We have recorded TF-$\mu$SR asymmetry time spectra for the VL of niobium for a large range 
of $(T,B_{\rm ext})$ values. From a global fit of the spectra with the Delrieu's
approximation for the form factor we deduce $v_{\rm F} = 2.0 \, (2) \times 10^5$~m/s,
in agreement with our previous estimate.\cite{Maisuradze13b}
From the measured $\tilde{a}$ and $\tilde{c}$ parameters and Eq.~\ref{Theory_51} 
the condensation energy $E_{\rm c}=N_0\Delta_0^2$ is determined. As an example, in 
Fig.~\ref{condensation_energy} 
we show the field dependence of 
$E_{\rm c} = -2\overline{B^Z}\tilde{a}/\mu_0\tilde{c}$ at $T=1.6$ K. The linear field 
dependence of $\Delta^2_0$ given by Eq.~\ref{Theory_4} is found to be a reasonable approximation. This is 
again consistent with the results of our analysis of spectra previously taken in the 
$B_{\rm c2}(T)$ vicinity. 

From the analysis of a asymmetry time spectrum recorded
near $T_{\rm c0}$ with the GL's theory we recall that
$\kappa_{\rm GL} = 0.89 \, (1)$.\cite{Maisuradze13a}

From a close look at Fig.~\ref{Measurement} we infer that the Delrieu's solution
for the form factor has a relatively large validity range for niobium, i.e.\ it is not only valid in
the immediate $B_{\rm c2}(T)$ vicinity but also for $ 0.6 \leq T \leq 2.5$~K with
$0.2 \, {\rm T}\leq B_{\rm ext} < B_{\rm c2}$. The lower temperature bound was  previously
given.\cite{Maisuradze13b} Since this solution also includes the Abrikosov's result and it is
numerically feasible to use it in a fit procedure, it should be seriously considered for the
analysis of TF-$\mu$SR data as a reliable alternative to a pure GL fit for clean s-wave
superconductors.\cite{Brandt97,Yaouanc97a}

\subsection{VL characteristic fields, field distributions and physical origins}
\label{Results_fields}

Having finished the analysis of the experimental data for various fields and temperatures with 
the Delrieu's theory, we now consider those from the Eilenberger's theory viewpoint. Hence, we 
can discuss the whole mixed phase, and not only the $B_{\rm c2}(T)$ vicinity. As already mentioned, 
Klein first calculated the detailed field profiles in the mixed state of niobium by solving the 
Eilenberger's equation.\cite{Klein87} Here based on a numerical algorithm explained in 
Refs.~[\onlinecite{ichioka97,ichioka99a,ichioka99b,Miranovic03,nakai06}], we have calculated $B^Z({\bf r})$ 
within a VL unit cell under periodic boundary conditions for various $B_{\rm ext}$ and $T$ appropriate 
for the present experimental situations. It will turn out later that $\kappa_{\rm GL}$=1.8 best 
describes the experimental data, thus all the following computations have been performed using this value. 
From this information a variety of physical quantities directly related to the present experiments can be 
deduced, that is, the field distribution ${\tilde D}_{\rm c}(B^Z)$ and therefore the characteristic three 
field values $B_{\rm min}$, $B_{\rm sad}$, and $B_{\rm vc}$ and their locations within a unit cell.

\begin{figure}
\includegraphics[scale=1.00]{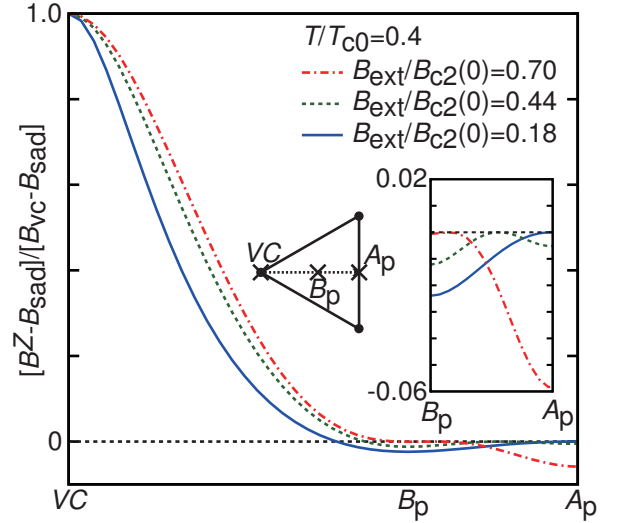}
\caption{(color online).
Results from the Eilenberger's theory for three field profiles along the vortex-core-$A_{\rm p}$  path, 
i.e.\ along $VC-A_{\rm p}$, via $B_{\rm p}$ in the direct space. The ratio value 
$B_{\rm ext}/B_{\rm c2}(0)=0.18$ corresponds to GL, $B_{\rm ext}/B_{\rm c2}(0)=0.44$ to KL, and  
$B_{\rm ext}/B_{\rm c2}(0)=0.70 $ to DL. We have chosen $T/T_{\rm c0} = 0.4$.
One of the insert recalls the definition of positions of interest in the unit cell and the other
insert magnifies the field profiles between $B_{\rm p}$ and $A_{\rm p}$. 
}
\label{profile}
\end{figure}

In order to understand the three possible field patterns, namely as predicted by Ginzburg-Landau (GL), 
Klein (KL), and Delrieu (DL) which we will identify through the analysis, we 
first show related field profiles for the three cases in Fig.~\ref {profile}. 
It is seen that

\noindent
(1) GL field profile for $B_{\rm ext}/B_{\rm c2}(0)=0.18$: $B_{\rm min}$ is located at $B_{\rm p}$ point and 
$B_{\rm sad}$ at $A_{\rm p}$ in the unit cell. The lowest edge of ${\tilde D}_{\rm c}(B^Z)$ occurs at $B_{\rm p}$.

\noindent
(2) DL field profile for $B_{\rm ext}/B_{\rm c2}(0)=0.70$: $B_{\rm min}$ is located at $A_{\rm p}$ point 
while $B_{\rm sad}$ is located at $B_{\rm p}$ point in 
the unit cell. The lowest edge of ${\tilde D}_{\rm c}(B^Z)$ occurs at $A_{\rm p}$.

\noindent
(3) KL field profile for $B_{\rm ext}/B_{\rm c2}(0)=0.44$: $A_{\rm p}$ is not saddle any more, but a local 
minimum. $B_{\rm p}$ is at the absolute minimum where the lowest edge of ${\tilde D}_{\rm c}(B^Z)$ occurs.
The saddle points are located in between $A_{\rm p}$ and $B_{\rm p}$. Those features are also  
seen from Fig. 17 in Ref.~\onlinecite{Klein87} where the contour plots for the three types of
distributions are displayed.

\begin{figure}
\includegraphics[scale=1.00]{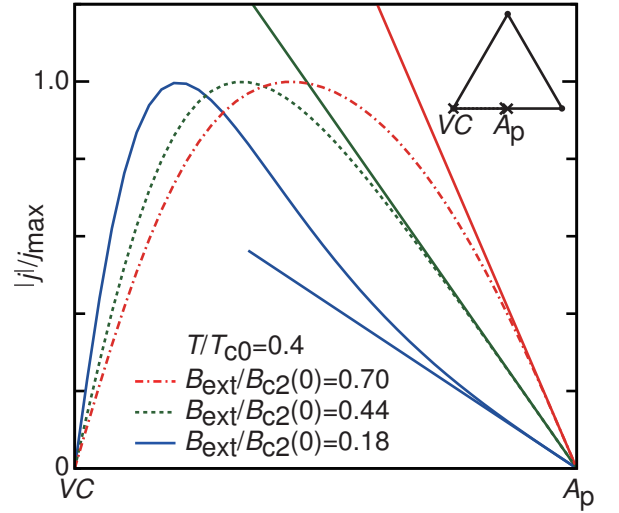}
\caption{(color online).
Three current profiles along the $VC-A_{\rm p}$ path  in the direct space. The parameters are the 
same as in Fig.~\ref{profile}. The currents are normalized to their maximum values. Note the changes
of the slope at $A_{\rm p}$ and of the maximum positions for the three profiles.
}
\label{j-profile}
\end{figure}

Let us now discuss the physical origins of those three kinds of field profiles. In 
Fig.~\ref{j-profile} 
we show the current profiles around the vortex core along the path from $VC - A_{\rm p}$ where each 
curve corresponds to that in Fig.~\ref{profile}. In GL ($B_{\rm ext}/B_{\rm c2}(0)=0.18$) the current maximum 
appears relatively near $VC$ and its amplitude quickly decays towards $A_{\rm p}$. Thus the current 
curve approaches $A_{\rm p}$ from above to  its tangential slope there, implying that the 
neighboring vortex cores are far apart and the vortex cores are not overlapped. This means that the 
farthest $B_{\rm p}$ point from the neighboring vortex cores in a unit cell is a $B_{\rm min}$
location. In contrast, the DL case shows that the current maximum moves towards $A_{\rm p}$, with
the tangent of the current amplitude being largest among the three profiles. This means that the 
neighboring vortex cores are densely packed with the vortex cores overlapped, causing $B_{\rm p}$ not to
be the minimum field location in a unit cell. The field profile is quite different from that in GL,
making $A_{\rm p}$ the minimum field location. In the KL limit those features are in between the GL and 
DL cases.

\begin{figure}
\includegraphics[scale=1.00]{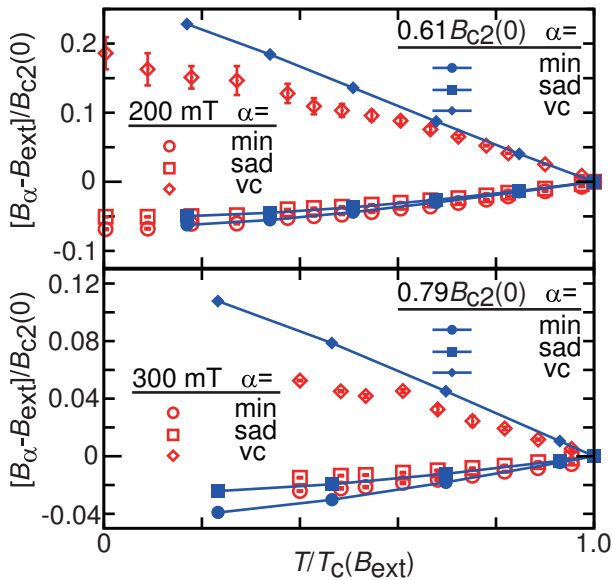}
\caption{(color online).
The three characteristic fields $B_{\alpha}$ ($\alpha =$~minimum, saddle and vortex core) versus the
normalized temperature $T/T_{\rm c}(B_{\rm ext})$ for two values of $B_{\rm ext}$. The computed values 
are shown by filled symbols linked by line segments. We have measured $T_{\rm c} (B_{\rm ext}) = 5.6$~K 
and $T_{\rm c} (B_{\rm ext}) = 3.95$~K for $B_{\rm ext} = 200$ and $300$~mT, respectively. }
\label{fig_Temperature_dependence}
\end{figure}

Figure~\ref{fig_Temperature_dependence} 
shows the comparison between the theoretical calculations and the experimental data of the 
$T$-dependence of $B_{\rm min}$, $B_{\rm sad}$ and $B_{\rm vc}$ at two 
$B_{\rm ext}$ values. The theoretical calculation has been done by varying the $\kappa_{\rm GL}$ value 
to best fit those three values near $T/T_{\rm c}(B_{\rm ext}) =1$. It turns out that the best fitting is achieved for 
$\kappa_{\rm GL}$=1.8. This value is twice as large as the nominal value 
$\kappa_{\rm GL} =0.89 \, (1)$ of the present sample mentioned before. We notice that the three types 
of field distributions, namely for the GL, KL and DL cases, are always present irrespective of the 
choice of $\kappa_{\rm GL}$. It is seen from Fig.~\ref{fig_Temperature_dependence} that 

\noindent
(1) The initial slopes of the three characteristic fields near $T/T_{\rm c}(B_{\rm ext}) =1$ are nicely reproduced 
for the two $B_{\rm ext}$ values.

\noindent
(2) Those  nice fittings continue to lower temperatures for $B_{\rm min}$ and $B_{\rm sad}$.

\noindent
(3) In contrast, $B_{\rm vc}$ starts to deviate towards lower temperatures. While the theoretical 
curves keep increasing linearly with large slopes, the temperature dependence of the experimental 
data is much weaker.

According to a previous calculation $B_{\rm vc}$ is expected to keep increasing towards zero 
temperature in the clean limit \cite{Miranovic03} because of the so-called Kramer-Pesch 
effect.\cite{Pesch74} As already noticed, a previous NMR experiment 
by Kung \cite{Kung70} on vanadium shows the expected linear temperature dependence of $B_{\rm vc}$ 
in a large temperature range towards zero temperature. Clearly it would be of much interest to 
perform TF-$\mu$SR measurements on a very clean vanadium sample to confirm Kung's result,
and to extend it to very low temperature.

\begin{figure}
\includegraphics[scale=1.00]{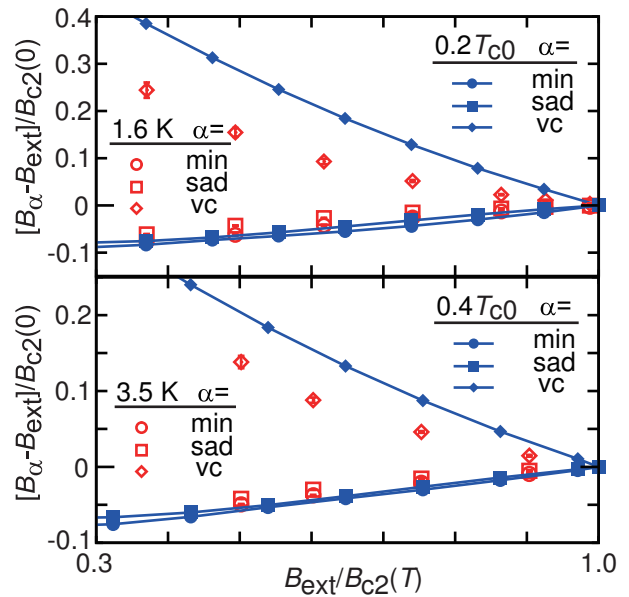}
\caption{(color online).
The three characteristic fields $B_{\alpha}$ ($\alpha =$~minimum, saddle, and vortex core) versus the
normalized field $B_{\rm ext}/B_{\rm c2}(T)$ at two temperatures. The computed values are shown 
by the filled symbols linked by line segments. Experimentally, $B_{\rm c2} = 405$ and $321$~mT 
for $T = 1.6$ and $3.5$~K, respectively. 
}
\label{fig_Field_dependence}
\end{figure}

We show the field dependences of the three characteristic fields in Fig.~\ref{fig_Field_dependence}.
It is seen that the theoretical predictions for $B_{\rm min}$ and $B_{\rm sad}$ follow nicely
the experimental results, and the qualitative field dependence of $B_{\rm vc}$ is explained, but 
quantitatively deviates because in those low temperatures the Kramer-Pesch effect is partially 
suppressed as mentioned above. Since $B_{\rm vc}$ reflects the spatial structure around a vortex core,
the partially suppressed Kramer-Pesch effect implies that the conical shape structure of $B^Z({\bf r})$
at the vortex core position is rounded relative to theoretical expectation. However, the linear field
tail of the distribution at high field, a  signature of the conical feature, is nicely observed at low 
temperature.\cite{Maisuradze13b}

\subsection{VL field map and contour plot}
\label{Results_map}

We have previously established the Delrieu's approximation to be reliable for a large part
of the mixed phase.
\begin{figure}
\includegraphics[scale=0.67]{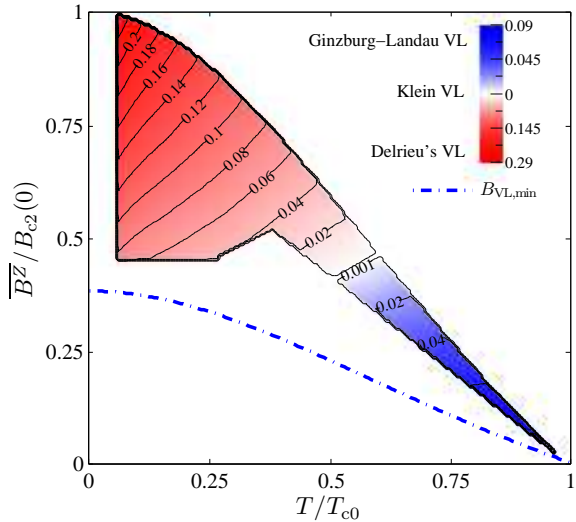}
\caption{(color online).
Contour map of $\delta B^{\rm n}_{\rm sad,min}$ for niobium with
${\bf B}_{\rm ext} \parallel \langle 111 \rangle$ computed with the Delrieu's solution for the
form factor. We have restricted the map to the temperature and field regimes properly described
by that solution. We specify the regimes consistent with Delrieu, Klein and Ginzburg-Landau.
The dashed-dotted line displays $B_{\rm VL,min}(T)$.
}
\label{Map_Delrieu}
\end{figure}
The computed $\delta B^{\rm n}_{\rm sad,min}$ map displayed in Fig.~\ref{Map_Delrieu} visualizes
the region of the mixed phase where the approximation is reliable for the analysis of our niobium 
data. Remarkably, the $(T, B_{\rm ext})$ points close to $B_{\rm c2}$ at which
$\delta B^{\rm n}_{\rm sad,min} =0$ effectively correspond in Fig.~\ref{fig_scan1} to the
temperature where the measured $\delta B^{\rm n}_{\rm sad,min}$ is minimum. This occurs
around $5.5$~K. It is because of the Gaussian smearing that
$\delta B^{\rm n}_{\rm sad,min}$ does not vanish experimentally.

From Delrieu, $\delta B^{\rm n}_{\rm sad,min} =0$ at the border between the DL and GL
VL region. This border belongs to the Klein's regime that we discuss now.

\begin{figure}
\includegraphics[scale=1.00]{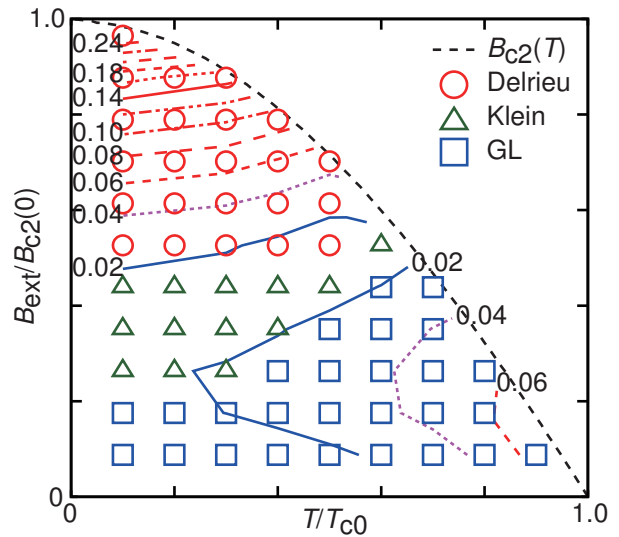}
\caption{(color online).
Contour map of $\delta B^{\rm n}_{\rm sad,min}$ for niobium with ${\bf B}_{\rm ext} \parallel \langle 111 \rangle$ 
computed with the Eilenberger's equation. In the Klein's region two minimal field positions exist; see
the insert on the right in Fig.~\ref {profile}. The data in Fig.~\ref{fig_scan2} (Fig.~\ref{fig_scan3}) 
correspond to scanning vertically at low (high) temperature. For the data in Fig.~\ref{fig_scan1} scanning 
is done parallel to the $B_{\rm c2}(T)$ line. 
}
\label{phase}
\end{figure}

In order to examine the diagram obtained experimentally in Fig.~\ref{Map_Delrieu},
we have done extensive computations to construct the corresponding diagram whose results
are displayed in Fig.~\ref{phase}. The overall features in Figs.~\ref{Map_Delrieu}
and \ref{phase} coincide, namely, DL occupies the higher field region while GL 
is located at lower field. The KL region is in the middle. However, the KL regime is no longer limited
to the border between the DL and GL regions: it has an appreciable extension.
This is a key result obtained from the Eilenberger's solution. The Delrieu's approximation is too rough
to capture the subtilities in $B^Z({\bf r})$ in the KL regime; see Fig.~\ref{profile}. In addition,
while in the vicinity of $B_{\rm c2}(T)$ the values of $\delta B^{\rm n}_{{\rm sad},{\min}}$ in 
Figs.~\ref{Map_Delrieu} and \ref{phase} strikingly correspond for a given field and temperature, outside
that regime deviations between the results in the two figures are clearly found: a given
$\delta B^{\rm n}_{{\rm sad},{\min}}$ value in the DL region is only weakly temperature dependent in 
Fig.~\ref{phase}, in contrast to the results presented in Fig.~\ref{Map_Delrieu}. Before going 
further in comparing the experimentally deduced and the theoretical diagram, some comments 
are in order.

Figure~\ref{Map_Delrieu} considers $\overline{B^Z}$ rather than $B_{\rm ext}$ because it is this 
parameter which enters into the Delrieu's theory. However, in pratice $\overline{B^Z} \simeq B_{\rm ext}$.
Because the lower field border of the measured phase is dependent on the experimental conditions, for the
sake of completeness, we have extended the theoretical diagram in Fig.~\ref{phase} to lower fields and 
temperatures than in Fig.~\ref{Map_Delrieu}. We stress that while the Delrieu's formula does not describe
the low temperature region of the mixed phase as seen in Fig.~\ref{Map_Delrieu}, the numerical solution 
of the Eilenberger's equation is expected to provide a proper description.

\begin{figure}
\includegraphics[scale=1.00]{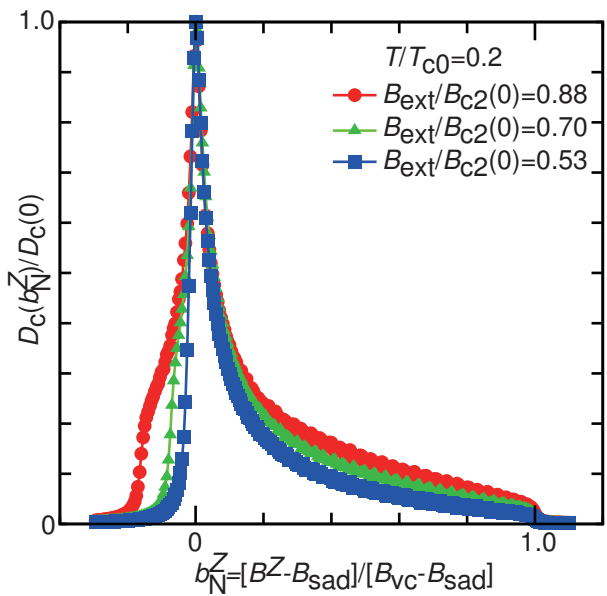}
\caption{(color online).
The normalized field distributions $D_{\rm c}(b^Z_{\rm N})/D_{\rm c}(0)$ for three field values at 
$T/T_{\rm c0} = 0.2$ corresponding to the data in the right hand panel of Fig.~\ref{fig_scan2}.
Those are traversing the map in Fig.~\ref{phase} vertically.
}
\label{fd}
\end{figure}

The data set shown in Fig.~\ref{fig_scan2} corresponds to scanning the field at a fixed 
low temperature in Fig.~\ref{phase}.  
As this map implies, $\delta B^{\rm n}_{\rm sad,min}$ increases as $B_{\rm ext}$ increases. In fact, 
as seen from Fig.~\ref{fd} 
where we show the computed normalized field distributions $D_{\rm c}(b^Z_{\rm N})/D_{\rm c}(0)$ under 
a fixed temperature ($T/T_{\rm c0} = 0.2$) for three field values, as $B_{\rm ext}$ increases, 
$B_{\rm min}$ moves to the lower field side, i.e.\ to the left of the field scale. 
This explains the fact in Fig.~\ref{fig_scan2} that 
$B_{\rm min}$ increases in absolute value as $B_{\rm ext}$ increases.

As for Fig.~\ref{fig_scan1} where the scanning path is taken parallel to the $B_{\rm c2}(T)$ line,
it is seen from Fig.~\ref{phase} that $\delta B^{\rm n}_{\rm sad,min}$ decreases first and
then increases towards lower temperatures, coinciding with the data in Fig.~\ref{fig_scan1}.

Finally as for GL,  Fig.~\ref{fig_scan3} shows that the two distributions are hardly distinguishable
because the two distributions are both inside the GL region in Fig.~\ref{phase} where the GL
distribution is universal and scaled.

Those various scanning data throughout the $B_{\rm ext} - T$ plane demonstrate precise correspondence
between Eilenberger's theory and experiment, supporting the existence of the three distinct 
characteristic field distributions, i.e.\ the GL, KL and DL distributions.

\section{Summary of the results obtained in this study; possible improvements of the analysis and
data recording}
\label{Summary_study}

In summary, combining TF-$\mu$SR measurements analyzed with the Delrieu's analytical solution
for the form factor --- supplemented with conventional phenomenological formulas for the
physical parameters --- and  the numerical solutions of the quasiclassical Eilenberger's equation
to get $B^Z({\bf r})$, we have established that the VL of niobium with ${\bf B }_{\rm ext}$ applied along
a three-fold axis is characterized by three successive regions as the sample is cooled down from
$T_{\rm c0}$. Hence, our work supports the predictions of Abrikosov, Klein, and Delrieu, respectively.
It seems that it has never been done previously.

The experimental data, notably the three regions in the mixed phase, are explained by the Eilenberger's theory taking 
the Fermi surface cylindrical, but the field and temperature dependences of $B_{\rm vc}$. Disturbing, 
a $\kappa_{\rm GL}$ parameter twice as large as the measured value has to be assumed. We know of two 
sources for possible explanations of the discrepancies.   

\noindent
(1) Our numerical solution of the Eilenberger's equation does not take into account the Fermi velocity 
anisotropy and gap anisotropy 
known to exist as seen in the $B_{\rm c2}$ anisotropy.\cite{Williamson70a,Williamson70b}
In particular the Fermi velocity anisotropy generally increases $B_{\rm c2}$ value,
thus causing the estimate of $\kappa_{\rm GL}$ to change.

\noindent
(2) We are regarding $\kappa_{\rm GL}$ as an effective parameter because the theory assumes the 
clean limit. Although our sample is extremely clean,\cite{Maisuradze13b} it is known that 
defects and impurities act to increase $\kappa_{\rm GL}$ from the nominal value.

We have to deal with three sources of field distributions at the muon site: the nuclear $^{93}$Nb
magnetic moments, the VL itself and the effect of the VL disorder. To a good approximation, the
component field distribution from the nuclear moments in a TF-$\mu$SR experiment is
Gaussian.\cite{Yaouanc11} We have just discussed how the description of the distribution from the VL 
itself can be improved. It is known that modeling the effect of the VL disorder with a Gaussian
field function as done in this report is a rough approximation. A close look at Fig.~2 of
Ref.~\onlinecite{Maisuradze13b} shows it definitively, in particular in the vicinity of the
low-field tail. In fact, the translational correlations of the vortex cores are neglected in the 
Gaussian approximation.\cite{Yaouanc13a} To progress we need to recognize that the VL is not a 
two-dimensional lattice, but a three-dimensional lattice, i.e.\ we are dealing with the flux-line 
lattice (FLL). As for any lattice, disorder has to be considered. In the FLL case we need to remember 
that the collective behaviour matters.\cite{Feigelman89} In the absence of dislocations, if disorder 
is not too strong the FLL is periodic, as clearly demonstrated by SANS measurements, but the FLL 
translational order decays only algebraically rather than exponentially,\cite{Klein01,Laver08} as expected 
theoretically.\cite{Nattermann90} In fact, a so-called Bragg glass state is expected,\cite{Giamarchi95} 
and observed.\cite{Klein01,Laver08} However, we stress that it was found for samples with appreciable 
disorder. It is still a challenge to observe it for a clean sample such as ours. A numerical method to 
account for the Bragg-glass state has been devised for the analysis of SANS measurements.\cite{Laver08} 
This has yet to be done for the $\mu$SR counterpart.

Neither the Delrieu's analytical solution, nor the numerical solution of the Eilenberger's equation
describes the measured distributions below 0.6~K.\cite{Maisuradze13b} A proper account of the
vortex-lattice residual disorder may round up the predicted
sharp conical field shape at the VL vortex cores and explain the measurements below $0.6$~K.

Up to now we have discussed possible improvements of the data analysis. But the experimental
conditions themselves could also be optimized.
All the TF-$\mu$SR asymmetry time spectra have been recorded on a single crystal disk
with ${\bf B}_{\rm ext}$ parallel to the disk axis; see pictogram in
Fig.~\ref{Measurement}. In this geometry inhomogeneities due to the demagnetization field near the sample boundaries
may have to be considered. An improved experimental setup would
require ${\bf B}_{\rm ext}$ to be applied perpendicular to the disk axis. However, for
the needed high $B_{\rm ext}$ values this is not possible since the positive muon is
a charge particle, and as such would be deflected from its
trajectory according to the Lorentz force. Hence, ${\bf B}_{\rm ext}$ should be kept
into the direction we have chosen. Therefore to improve on the experimental conditions,
an elliptical single sample would have to be used.

The form factors of a vortex lattice can be studied by the SANS technique. It is well known that the exact 
solution of GL theory gives some $K_{m,n}$ form factors of opposite sign relative to those predicted by analytical 
approximations of GL theory or the London model; See Ref.~\onlinecite{Brandt97} for a discussion.
Regarding Delrieu's solution, the signs of the form factors are the same 
as is in the Abrikosov's solution and are given by the factor $(-1)^{mh}$.\cite{Maisuradze13a}
Only the magnitude of the form factors varies with the values of the parameters
$\tilde{b}$ and $\tilde{c}$. The signs of the form factors at high order from the Eilenberger 
solution have still to be evaluated. This requires to get the solution accurate enough
to extract those higher order harmonics because those become extremely small numbers.

\section{Conclusions and perspectives}
\label{Conclusions}

In conclusion, combining the TF-$\mu$SR experimental technique with the Delrieu's analytical 
solution and numerical solutions of the quasiclassical Eilenberger's equation, we have observed
the theoretically expected three regions in the mixed phase of niobium  with ${\bf B }_{\rm ext}$ 
applied along a three-fold axis. We do not know of any previous experimental observation of the 
three regions. Our results should apply to any clean s-wave superconductor with a triangular vortex 
lattice. 

The experimental data have been recorded at high statistics and the analysis has been done with
advances methods. Possible improvements of the data analysis and experimental conditions have been
pointed out. We hope that our work will motivate people to analyze TF-$\mu$SR asymmetry time spectra
for other s-wave superconductors with the framework presented here. An obvious candidate
is vanadium, the sample of which should be in the extremely clean limit.

\section*{Acknowledgments}

We thank P. Dalmas de R\'eotier for a careful reading of the manuscript.
The $\mu$SR measurements were performed at Swiss Muon Source (S$\mu$S), Paul Scherrer 
Institute (PSI), Villigen, Switzerland. We acknowledge partial support from NCCR MaNEP 
sponsored by the Swiss National Science Foundation. K. M. was supported by JSPS 
(Grant Nos. 21340315 and 9134919315) and "Topological Quantum Phenomena'' 
(No. 2510371615) KAKENHI on Innovation Areas from MEXT.

\bibliography{reference_Nb.bib}

\end{document}